\documentclass[conference]{IEEEtran}
\IEEEoverridecommandlockouts
\usepackage{cite}
\usepackage{amsmath,amssymb,amsfonts}
\usepackage{algorithmic}
\usepackage{graphicx}
\usepackage{textcomp}
\usepackage{braket}
\usepackage{xcolor}
\usepackage{soul}
\def\BibTeX{{\rm B\kern-.05em{\sc i\kern-.025em b}\kern-.08em
    T\kern-.1667em\lower.7ex\hbox{E}\kern-.125emX}}

\begin{document}

\title{Efficient re-sampling in quasi-probability decompositions\\

\thanks{This work was supported by the Swiss National Science Foundation through Projects No. 200020\_215172, 200021-227992, and 20QU-1\_215928, and as part of NCCR SPIN (grant number 225153).}
}

\author{
\IEEEauthorblockN{
Sara Santos\textsuperscript{1,2},
Stefan Woerner\textsuperscript{3},
Vincenzo Savona\textsuperscript{1,2},
Julien Gacon\textsuperscript{3}
}
\IEEEauthorblockA{
\textsuperscript{1}\textit{Institute of Physics, École Polytechnique Fédérale de Lausanne (EPFL), Lausanne, Switzerland} \\
\textsuperscript{2}\textit{Center for Quantum Science and Engineering, École Polytechnique Fédérale de Lausanne (EPFL), Lausanne, Switzerland} \\
\textsuperscript{3}\textit{IBM Research, R\"uschlikon, Switzerland} \\
\texttt{sara.alvesdossantos@epfl.ch}, \texttt{wor@zurich.ibm.com}, \\
\texttt{vincenzo.savona@epfl.ch}, \texttt{jul@zurich.ibm.com}
}
}
\maketitle

\begin{abstract}
Near-term quantum devices are limited by noise and hardware constraints, motivating algorithmic approaches that trade circuit complexity for increased sampling overhead.
Quasi-probability decompositions (QPDs), for example, allow replacing non-local operations by multiple circuits with local operations, but the associated sampling overhead generally scales exponentially and limits their practicality.
In this work, we introduce a reweighting strategy for QPDs for circuits with the same variational structure across parameter settings, reusing samples and thereby reducing the sampling overhead.
We first demonstrate this approach by estimating fidelities between parameterized quantum states, a key primitive in variational time evolution and quantum kernel methods. Importantly, this setup allows controlling the exponential QPD sampling overhead while preserving the structure of the state-encoding ansatz.
We then apply the method to estimate the real part of the quantum geometric tensor using the simultaneous perturbation stochastic approximation and find that, 
in the presence of realistic hardware noise, our method outperforms other standard estimation techniques.
These results highlight the potential of reweighting strategies to extend the applicability of QPD-based methods in variational quantum algorithms.
\end{abstract}


\section{Introduction}
   Quantum computing is an emerging field with the potential to impact a wide range of applications, from simulating quantum systems~\cite{Feynman_book_1982,Maskara_molecules_simulation_2025,Ollitrault_molecular_dynamics_2020,Lloyd_universal_quantum_sim_1996,Barkoutsos_electronic_structure_calculations_2018} to solving hard classical tasks~\cite{Farhi_quantum_adiabatic_algo_2001,Zhou_QAOA_2020,Harrigan_QAOA_2021,Jordan_DQI_2025}. Since the early proposals of quantum simulation by~\cite{Feynman_book_1982} and~\cite{Lloyd_universal_quantum_sim_1996}, substantial progress has been made in physical realizations of quantum computers~\cite{Bluvstein_atom_arrays_2022,Bluvstein_rydberg_atoms_2024,Philips_universal_control_spin_qubits_2022,Cao_66qubit_entanglement_2023,VanDamme_CMOS_qubit_fabrication_2024,Pino_trapped_ion_2021}, with demonstrations scaling from few-qubit experiments~\cite{Sackett_entanglement_ions_2000,SchmidtKaler_CNOT_2003,Gulde_DeutschJozsa_on_trapped_ions_2003,Monroe_scaling_trapped_ions_2013} to the 100-qubit regime~\cite{Arute_supremacy_2019,Shaydulin_QAOAadvantage_2024}.
   
    Despite this progress, hardware noise and restricted device connectivity limit the complexity of circuits that can be reliably executed on near-term devices. In particular, implementing long-range interactions on fixed-connectivity hardware requires additional SWAP operations, leading to deeper circuits and increased error. Designing hardware-efficient algorithms that respect the device's constraints is therefore a central challenge.
    
    Circuit-knitting techniques can address these limitations by decomposing large circuits into smaller subcircuits that are  more suitable for available hardware, and are recombined via classical post-processing~\cite{Mitarai_Fujii_QPD_2021,Piveteau_Sutter_CircuitKnittingClassicalCommunication_2022}. Among these, quasi-probability decomposition methods (QPDs) allow expressing non-local operations as signed mixtures of local, device-native ones, and extends naturally to error mitigation applications~\cite{Temme_PEC_2017,Piveteau_QPD_2022,CarreraVazquez_SWAP_QPD_2024}. A limitation of QPDs is their sampling overhead, which generally grows exponentially with the number of decomposed operations. In practice, however, for weakly entangling gates, the quasi-probability distribution is often dominated by a few terms, and can be efficiently approximated via Monte Carlo sampling.
    
    In this work, we introduce a reweighting framework for QPDs that can improve sampling efficiency when evaluating the same observable across a family of parameterized variational circuits. Instead of constructing separate decompositions, we sample from a single reference QPD and reuse these samples via self-normalized importance sampling to approximate other parameter settings. This is particularly efficient when the parameters differ only slightly from the reference, as can occur in various variational algorithms.
    
    As an application, we estimate fidelities between parameterized quantum states, a fundamental subroutine in variational algorithms such as variational quantum simulation~\cite{Yuan_TheoryVQS_2019,Motta_QITE_2020,Endo_VQSGeneralProcesses_2020,Lin_CompressedTimeEvol_2021,Barison_pVQD_2021,Gacon_saqite_2023,Gacon_dualqite_2024}, and quantum natural gradient methods~\cite{Stokes_QNG_2020,Wierichs_NatGradVQE_2020,Gacon_spsa_2021}. This setting is particularly well suited to our approach, as fidelities are often evaluated between closely related states, enabling the efficient reuse of QPD samples. In contrast to standard methods such as the Hadamard test~\cite{Kitaev_hadamard_test_1995,NielsenChuang_QC_2010} or compute–uncompute~\cite{Gacon_dualqite_2024}, our method does not require increased circuit depth or ancilla qubits.
    
    Using this fidelity estimation as a subroutine, we estimate the real part of the quantum geometric tensor (QGT) via simultaneous perturbation stochastic approximation (SPSA), which requires multiple fidelity evaluations at nearby parameter values. By reusing samples from a single reference QPD, these fidelities can be obtained without additional sampling, reducing the measurement overhead. Furthermore, our numerical simulations in noisy regimes show that, due to the reduced circuit depth, the QPD-based approach can achieve lower errors than other fidelity estimation methods.
    
    The article is structured as follows. In Section~\ref{sec:qpd_theory}, we review quasi-probability decompositions, introduce a controlled-R$_Z$ decomposition, and present fidelity estimation via a QPD of the Hadamard test. In Section~\ref{sec:reweighting}, we show how sampled circuits can be reused via reweighting, discussing the necessary conditions, and describe how the QPD-fidelity estimator can be used to efficiently compute the QGT via SPSA, before we conclude in 
    Section~\ref{sec:conclusions}.
    
    \section{Quasi-probability decomposition}
    \label{sec:qpd_theory}
    
    Consider a quantum circuit that is difficult to realize in practice, since it may contain gates that are non-local and whose implementation
    would require routing overhead. Such a circuit can be executed
    using a set of easier-to-implement operations via a QPD~\cite{Mitarai_Fujii_QPD_2021}, at the cost of increased sampling overhead.
    
    To describe quantum circuits in a form suitable for QPDs, it is convenient to work in the superoperator representation.
    Let $U$ denote a quantum circuit acting on an initial state $\boldsymbol{\rho}$, with output $\mathcal{U} := U \boldsymbol{\rho} \,U^{\dagger}$.
    
    A QPD expresses $\mathcal{U}$ as a linear combination of implementable channels $u_{i}$~\cite{Mitarai_Fujii_QPD_2021},
    \begin{equation}
        \label{eq:qpd_decomposition}
        \mathcal{U} = \sum_{i \in \xi}a_{i} \, u_{i} \, ,
    \end{equation}
    where $a_{i} \in \mathbb{R}$, $\sum_{i} a_{i} = 1$, and $\xi$ is an index set. Since $a_i$ may be negative, this is a \emph{quasi-probability} decomposition. A QPD thus defines a (generally overcomplete) basis for the space of superoperators, and a quantum channel may admit multiple distinct decompositions. These can be adapted to the target hardware, with the aim that each $u_{i}$ is simpler to implement than $\mathcal{U}$. Note that the channels $u_{i}$ need not be unitary and may include noisy operations or measurements.
    
    In practice, \eqref{eq:qpd_decomposition} is estimated via Monte Carlo by sampling channels $u_{i}$ according to probabilities $p_i$,
    \begin{equation}
        p_{i} = \frac{|a_{i}|}{\gamma}, \qquad \gamma = \sum_{i\in\xi} |a_{i}| \, ,
    \end{equation}
    and constructing the estimator
    \begin{equation}
    \label{eq:qpd_estimator}
        \hat{\mathcal{U}}= \frac{\gamma}{M}\sum_{m=1}^{M} \mathrm{sgn}(a_{m}) \, u_{m} 
        = \gamma \sum_{i\in\xi}\hat{p}_{i}\mathrm{sgn}(a_{i}) u_{i}\,,
    \end{equation}
    where $\hat{p}_{i}=\hat{n}_{i}/M$, with $\hat{n}_{i}$ the number of times the
    channel $u_{i}$ is sampled. 

    In this framework, the expectation value of a projective measurement described by a Hermitian operator $O$ performed after $U$, can be obtained from the QPD estimator defined in \eqref{eq:qpd_estimator} together with the linearity of the trace $\mathrm{tr}(\cdot)$:
    \begin{equation}
        \mathrm{tr}\!\left( O\,\hat{\mathcal{U}} \right)
        =
        \sum_{i \in \xi} a_i \,
        \mathrm{tr}( O\,u_i)\,,
    \end{equation}
    which is an unbiased estimator of $\mathrm{tr}( O\,\mathcal{U})$
    and has variance $\mathcal{O}(\gamma^{2} / M)$, where $\gamma \geq 1$ quantifies the sampling overhead~\cite{Piveteau_QPD_2022}.
    
    This formalism extends to a sequence of ${n_c}$ channels, 
    $\mathcal{U}= \mathcal{U}_{n_c}\circ \cdots \circ \mathcal{U}_{1}$, by combining the individual QPDs of each channel. In this case, the coefficients and the sampling overhead simply factorize, with a total
    \begin{equation}
        \gamma= \prod_{j=1}^{n_c} \gamma_{j}\,.
    \end{equation}
    Consequently, the QPD set of operations and the sampling overhead grow exponentially with the number of decomposed, or \emph{cut}, gates $n_c$. For this reason, practical applications must be restricted to regimes where the QPD remains efficient. In particular, for weakly entangling gates, such as two-qubit rotation gates with small angles, the quasi-probability distribution is concentrated on a few dominant terms, resulting in a small $\gamma$-factor. 
    In the following, we discuss the controlled-R$_Z$ rotation as an example.

    \subsection{Decomposition of the controlled-R$_Z$ gate}
    \label{sec:qpd_crz}
    
    As an illustrative example, we present a QPD decomposition of the controlled-R$_Z$
    rotation (CR$_Z$) gate on two qubits. In what follows, we work in the superoperator
    representation and denote by $\mathcal{S}(U)$ the superoperator associated with
    a unitary $U$.
    
    The CR$_Z$ gate acts as the identity on the target qubit when the control qubit
    is in the state $\ket{0}$, and applies a single-qubit $Z$ rotation when the control
    is in the state $\ket{1}$. It can be written as
    \begin{equation}
        \label{eq:crz_definition}
        \mathrm{CR}_Z(\theta) = \ket{0}\!\bra{0}\otimes \mathbb{I}
        + \ket{1}\!\bra{1}\otimes \mathrm{R}_Z(\theta)\,,
    \end{equation}
    where $\mathrm{R}_Z(\theta)=\exp(-i\,\theta/2\,Z)$.
    
    While multiple methods exist to construct QPDs~\cite{Piveteau_QPD_2022,Temme_PEC_2017}, here we decompose the CR$_Z$
    gate using a known QPD applicable to two-qubit rotation gates~\cite{Mitarai_Fujii_QPD_2021}.
    We first express CR$_Z$ as a product of single- and two-qubit gates:
    \begin{equation}
        \label{eq:crz_transpilation}
        \mathrm{CR}_Z(\theta) = \mathrm{R}_{ZZ}\!\left(-\frac{\theta}{2}\right)
        \left[\mathbb{I}\otimes \mathrm{R}_Z\!\left(\frac{\theta}{2}\right)\right] \, ,
    \end{equation}
    where $\mathrm{R}_{ZZ}(\theta)=\exp\!{(-i\,\theta/2 \,Z\otimes Z)}$. Both these gates can be decomposed into operations that are independent of $\theta$, as shown
    in Appendix~\ref{app:crz_decomposition}. By combining the QPDs of the R$_{ZZ}$ and
    R$_Z$ gates, we obtain the following QPD of the CR$_Z$
    gate in superoperator form:
    \begin{equation}
        \label{eq:qpd_crz}
        \begin{aligned}
             & \mathcal{S}\left(\operatorname{CR}_Z(\theta)\right) = \cos^{4}\left(\frac{\theta}{4}\right) \mathcal{S}\left(\mathbb{I}\otimes \mathbb{I}\right)+ \sin^{4}\left(\frac{\theta}{4}\right) \mathcal{S}\left(Z \otimes \mathbb{I}\right) \\
             & \quad + \frac{1}{4}\sin^{2}\left(\frac{\theta}{2}\right) \left(\mathcal{S}\left(\mathbb{I}\otimes Z\right) + \mathcal{S}\left(Z \otimes Z\right)\right)                                                                             \\
             & \quad + \frac{1}{2}\sin\left(\frac{\theta}{2}\right)\cos^{2}\left(\frac{\theta}{4}\right) \left(\mathcal{S}\left(\mathbb{I}\otimes S\right) - \mathcal{S}\left(\mathbb{I}\otimes S^{\dagger}\right)\right)                          \\
             & \quad + \frac{1}{2}\sin\left(\frac{\theta}{2}\right)\sin^{2}\left(\frac{\theta}{4}\right) \left(\mathcal{S}\left(Z \otimes S^{\dagger}\right)- \mathcal{S}\left(Z \otimes S\right)\right)                                           \\
             & \quad + \sum_{\alpha=\pm 1}\frac{\alpha}{4}\sin(\theta) \left(\mathcal{M}_{\alpha}\otimes\mathcal{S}\left(S^{\dagger}\right)- \mathcal{M}_{\alpha}\otimes \mathcal{S}\left(S\right)\right)                              \\
             & \quad + \sum_{\alpha=\pm 1}\frac{\alpha}{2}\sin\left(\frac{\theta}{2}\right) \left(\mathcal{S}\left(S^{\dagger}\right)\otimes \mathcal{M}_{\alpha}- \mathcal{S}\left(S\right) \otimes \mathcal{M}_{\alpha}\right)       \\
             & \quad + \sum_{\alpha=\pm 1}\frac{\alpha}{2}\sin^{2}\left(\frac{\theta}{2}\right) \left(\mathcal{M}_{\alpha}\otimes \mathcal{S}\left(\mathbb{I}\right)- \mathcal{M}_{\alpha}\otimes \mathcal{S}\left(Z\right)\right) \,,
        \end{aligned}
    \end{equation}

    where $\mathcal{M}_{\alpha}$ denotes the (unnormalized) projection onto the $Z$-basis, postselected
    on outcome $\alpha\in\{+1,-1\}$, defined as
    \begin{equation}
        \mathcal{M}_{\alpha}= \left(\mathbb{I}+\alpha Z\right) \boldsymbol{\rho} \left(\mathbb{I}+\alpha Z\right) \,.
    \end{equation}
    
    The $\gamma$-factor of this decomposition is 
    \begin{equation}
        \label{eq:gamma_crz}
        \gamma_{\mathrm{CR}_Z}(\theta) = 1 + \left| \sin \left(\frac{\theta}{2}\right)\right| \left(2+ \left| \sin\left(\frac{\theta}{2}\right) \right| + \left| \cos\left(\frac{\theta}{2}\right)\right| \right)
        \,.
    \end{equation}
    This is larger than the known optimal $\gamma_{\mathrm{opt}}=1+2|\sin(\theta/2)|$ for QPDs of parameterized two-qubit gates, which can be achieved, for example, by only decomposing the $R_{ZZ}$ gate in~\eqref{eq:crz_transpilation}, while leaving the single-qubit $\mathrm{R}_Z$ gate unchanged~\cite{Mitarai_Fujii_QPD_2021}. 
    However, the resulting local channels retain an explicit angle dependence, 
    $u_i = u_i(\theta)$, which will turn out to violate conditions required for 
    reweighting techniques introduced in Section~\ref{sec:reweighting}. 
    Instead, we focus on a decomposition with angle-independent local channels $u_i \equiv \mathrm{const}$.
 
    \subsection{Application to fidelity estimation}
    \label{sec:application_fidelity}
    The fidelity quantifies the closeness between two quantum states 
    and, for pure states $\ket{\psi}$ and $\ket{\phi}$,  defined as
    \begin{equation}
    \label{eq:fidelity_real_imag}
    \left|\braket{\psi|\phi}\right|^{2}
    = \mathrm{Re}\left[\braket{\psi|\phi}\right]^{2} + \mathrm{Im}\left[\braket{\psi|\phi}\right]^{2}\, \in [0, 1],
    \end{equation}
    which equals $0$ for orthogonal states and $1$ when the states coincide up to a global phase. As discussed in the introduction, computing the fidelity is a central subroutine in many quantum algorithms and in the following we focus on its efficient estimation on a quantum device.

    Consider two states $\ket{\psi} = U\ket{0}$ and $\ket{\phi} = V\ket{0}$ prepared by unitary circuits $U$ and $V$. The fidelity can be written as 
    \begin{equation}
        \label{eq:cu_fidelity}
        |\!\braket{\psi|\phi}|^{2}
        = |\!\braket{0|U^\dagger V|0}|^{2}\,,
    \end{equation}
    which is the probability of measuring the all-zero state after applying $U^{\dagger}V$ to $\ket{0}$. This leads to the compute-uncompute (CU) method~\cite{Gacon_dualqite_2024}, which estimates the fidelity by measuring this probability. Although simple, this approach may result in a deep circuit $U^{\dagger}V$ and large errors on near-term devices.
    
    Beyond the CU method, several fidelity-estimation techniques reduce circuit depth. For example, the classical-shadows method~\cite{Huang_classical_shadows_2020} estimates the overlap from randomized measurements, but it is efficient mainly for highly structured states and otherwise incurs exponential measurement overhead. Another standard approach is the Hadamard test (HT)~\cite{Kitaev_hadamard_test_1995, NielsenChuang_QC_2010}, which computes the real and imaginary parts of the overlap using an ancilla qubit and controlled operations (see Fig.~\ref{fig:hadamard_test}). While this method requires controlled versions of the state-preparation unitaries, which can substantially increase circuit depth relative to the CU method, for states prepared by the same parameterized ansatz it can be compressed to the ansatz depth. As this setting commonly arises in variational algorithms, we focus on the HT in the following and show how QPD removes the need for an ancilla qubit.

    \subsubsection{Hadamard test for fidelity estimation}
    \label{sec:hadamard_test}

\begin{figure}[t]
        \centering
        \includegraphics[width=0.8\linewidth]{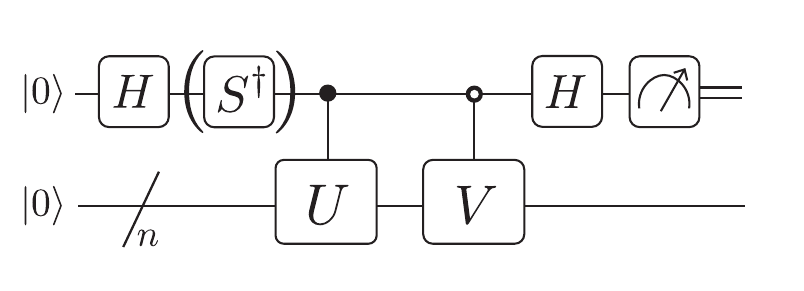}
        \caption{Hadamard test circuit for computing the real and imaginary
        parts of
        $\langle\phi|\psi\rangle=\langle 0|V^{\dagger}U|0 \rangle$. $H$ denotes the Hadamard gate and the imaginary part of the overlap is estimated with an additional $S^{\dagger}$ gate. The states are encoded via controlled implementations of the unitaries $U$ and $V$, with $V$ applied using an open control, i.e., the gate is activated when the control qubit is $\ket{0}$.}
        \label{fig:hadamard_test}
    \end{figure}

    Let $\ket{\psi} = U\ket{0}$ and
    $\ket{\phi} = V\ket{0}$ be two $n$-qubit states prepared by the unitaries $U$ and $V$, respectively.
    Fig.~\ref{fig:hadamard_test} illustrates the Hadamard test circuit.
    \begin{figure}[t]
        \centering
        \includegraphics[width=\linewidth]{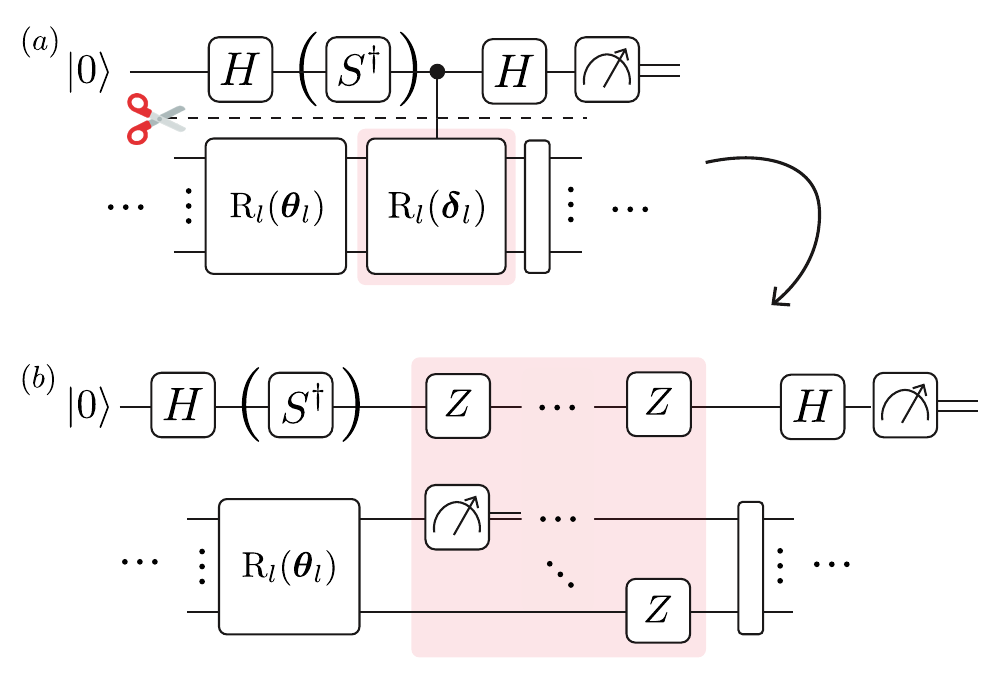}
        \caption{Decomposition of the compressed Hadamard test circuit: (a) each
        controlled gate is cut and replaced by a local channel from a predefined
        decomposition set; (b) the cuts separate the ancilla qubit from the
        remaining circuit, leaving only local channels in place of the controlled
        gates.}
        \label{fig:cut_compressed_hadamard_test}
    \end{figure}
    Immediately before the ancilla measurement, the real and imaginary circuits prepare
    \begin{equation}
        \begin{split}
            \ket{s_{\mathrm{re}}}&= \frac{\left(\ket{0}+\ket{1}\right)\ket{\phi}+\left(\ket{0}-\ket{1}\right)\ket{\psi}}{2}
            \,, \\
            \ket{s_{\mathrm{im}}}&= \frac{\left(\ket{0}+\ket{1}\right)\ket{\phi}-i\left(\ket{0}-\ket{1}\right)\ket{\psi}}{2}
            \,.
        \end{split}
    \end{equation}
    Measuring the ancilla yields
    \begin{equation}
            p_{0}^{{\mathrm{re}}}= \frac{1}{2}+\frac{\mathrm{Re}[\braket{\phi|\psi}]}{2}\,,
            \quad
            p_{0}^{{\mathrm{im}}}= \frac{1}{2}+\frac{\mathrm{Im}[\braket{\phi|\psi}]}{2}\,.
    \end{equation}
    
    Using $p^{\mathrm{re}}_{1}= 1-p^{\mathrm{re}}_{0}$, the real part $\hat{R}$ can be written as the expectation value of the ancilla Pauli-$Z$ operator,
    \begin{equation}
        \label{eq:real_fidelity_ht}
        \hat{R}= \langle Z\rangle= p^{{\mathrm{re}}}_{0}- p^{{\mathrm{re}}}_{1}\, ,
    \end{equation}
    with an analogous expression for the imaginary part $\hat{J}$. The fidelity estimator is then given by
    \begin{equation}
        \hat{F}=\hat{R}^2+\hat{J}^2\,.
    \end{equation}

    \subsubsection{Compressed Hadamard test and its decomposition}
    \label{sec:hadamard_qpd}
      
    When both states are described by the same parameterized ansatz, $U = U(\boldsymbol{\theta})$, with $\boldsymbol{\theta}\in\mathbb{R}^d$, the Hadamard test circuit can be simplified. Consider that $U$ is a sequence of layers R$_l(\boldsymbol{\theta}_l)$ of single-qubit rotations and layers $W_l$ of fixed, multi-qubit gates, that is  
    $U(\boldsymbol{\theta}) = \prod_{l}\,\mathrm{R}_{l}(\boldsymbol{\theta}_l)W_{l}$.
    Using that $\mathrm{R}_l(\boldsymbol{\theta}_l + \boldsymbol{\delta}) = \mathrm{R}_l(\boldsymbol{\theta}_l)\mathrm{R}_l(\boldsymbol{\delta})$, we can see that
    for the states $\ket{\psi(\boldsymbol{\theta})}=U(\boldsymbol{\theta})\ket{0}$ and $\ket{\psi(\boldsymbol{\theta}+\boldsymbol{\delta})}=U(\boldsymbol{\theta}+\boldsymbol{\delta})\ket{0}$
    the standard Hadamard test circuit is equivalent to a compressed version (see Fig.~\ref{fig:cut_compressed_hadamard_test}$\,$(a)) in which the ancilla-controlled gates depend only on the parameter difference $\boldsymbol{\delta}$. This reduces the number of controlled gates from $2d$ to $d$ and halves the number of two-qubit gates, lowering circuit depth and error rates.

    The compressed circuit can be further simplified using QPD-based circuit cutting (see Section~\ref{sec:qpd_theory}), which decomposes the controlled rotations into local operations and separates the ancilla from the state-preparation circuit, splitting the original circuit into two parts, which can be executed independently and recombined classically to estimate the overlap. Note that the cut circuit preserves the structure of the original ansatz $U$, with the same width and depth when counted in terms of multi-qubit gates. 

    Using the notation $\hat{R}$ and $\hat{J}$ introduced earlier, we obtain
    \begin{equation}
        \label{eq:fidelity_estimators}
        \hat{R}= \gamma\sum_{i\in \xi}\hat{p}_{i}
         \hat{{\sigma}}_{i}\langle Z\rangle_{i}\, ,
    \end{equation}
    with
    \begin{equation}
        \hat{{\sigma}}_{i}= \mathrm{sgn}(a_{i}) \sum_{s\in \mathcal{O}_i}\hat{q}_{s}
        \prod_{m\in s}(-1)^{\delta_{m,1}}\, ,
    \end{equation}
    where $\mathcal{O}_{i}$ denotes the set of mid-circuit measurement outcomes, represented as $\{0,1\}$ bitstrings, and $\hat{q}_{s}$ their frequencies. Note that $\hat{\sigma}_{i}$ encodes both the sign of the coefficient associated with the $i$-th decomposed circuit and the measurement outcome(s), which determine the implemented projection in the QPD: $\mathcal{M}_{\alpha=+1}$ for outcome $0$ and $\mathcal{M}_{\alpha=-1}$ for outcome $1$. The expectation values $\langle Z\rangle_{i}$ are computed on the ancilla qubit and can be evaluated classically (see Appendix~\ref{app:top_exp_value}).

    \subsubsection{Numerical validation}
       \begin{figure}[t]
        \centering
        \includegraphics[width=\linewidth]{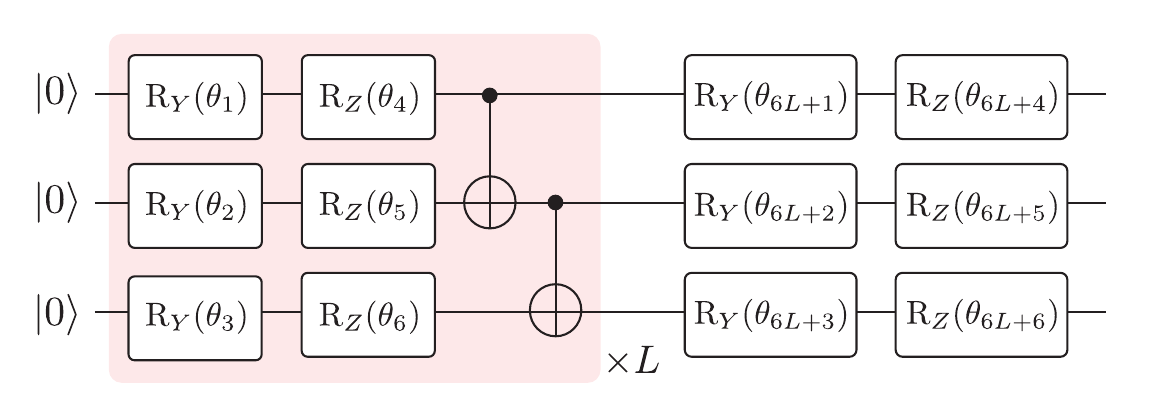}
        \caption{Schematic representation of a layered ansatz with 3 qubits. This
        ansatz is composed of blocks of R$_Z$ and R$_Y$ gates
        applied to all qubits, inter-layered by CNOT gates. $L$ represents
        the number of layers, i.e., the number of times the shaded block is
        repeated (with different parameters).}
        \label{fig:efficient_su2}
    \end{figure}
    \begin{figure}[ht]
        \centering
        \includegraphics[width=0.9\linewidth]{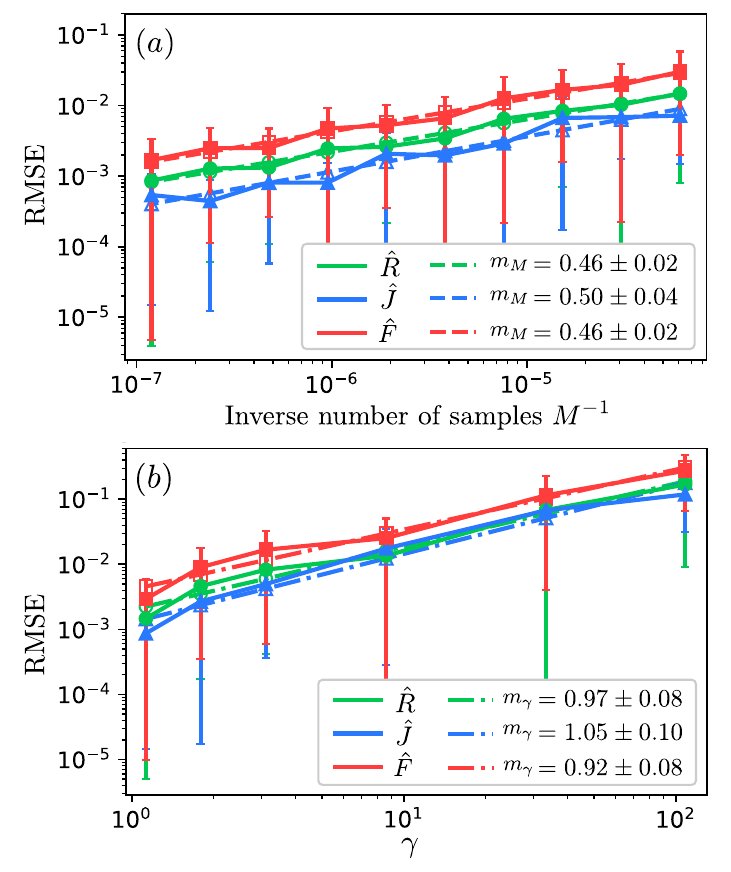}
        \caption{Numerical results for the RMSE of the fidelity estimator, averaged
        over 10 runs. (a) RMSE vs.~$M^{-1}$ with fixed $h = 0.1$. (b)
        RMSE vs.~$\gamma$ with fixed $M = 5\times10^{4}$.}
        \label{fig:rmse_fidelity}
    \end{figure}

    We estimate the fidelity between $\ket{\psi(\boldsymbol{\theta})}=U(\boldsymbol{\theta})\ket{0}$ and $\ket{\psi(\boldsymbol{\theta}+h\mathbf{1})}=U(\boldsymbol{\theta}+h\mathbf{1})\ket{0}$, where $U$ is a 3-qubit layered ansatz (see Fig.~\ref{fig:efficient_su2}) with $L=2$ alternating R$_Y$/R$_Z$ layers and CNOT gates; $\boldsymbol{\theta} \in \mathbb{R}^{d}$, $\mathbf{1}$ is the $d$-dimensional all-1 vector and $h \in \mathbb{R}$ controls the perturbation. In the compressed Hadamard test circuit, controlled-$\mathrm{R}_Z$ (CR$_Z$) and controlled-$\mathrm{R}_Y$ (CR$_Y$) gates appear. The latter are transpiled into CR$_Z$ and other fixed single-qubit gates, allowing us to apply the CR$_Z$ decomposition introduced in Section~\ref{sec:qpd_crz}.
    \vspace{-0.02in}
    We evaluate the root mean-squared error (RMSE) of the estimators for the real and imaginary parts of the overlap and for the fidelity as a function of the number of QPD samples $M$ and the overhead $\gamma$. Specifically, we consider (i) fixed $M=5\times10^{4}$ while varying $h$, and (ii) fixed $h=0.1$ while varying $M$. The results are presented in Fig.~\ref{fig:rmse_fidelity}.
    
    We observe that the RMSE of the fidelity scales as $\mathcal{O}(\gamma/\sqrt{M})$, consistent with standard error scaling analysis presented in Appendix~\ref{app:variance_fidelity_qpd}.
    \vspace{-0.02 in}
    
    \section{Sample-efficient QPD with reweighted distributions} \label{sec:reweighting}
      
    When parameterized circuits $U(\boldsymbol\theta)$ and $U(\boldsymbol\theta^\prime)$ differ only in gate angles, their QPDs can share the same set of channels $u_i$. In this case, samples drawn from a QPD for a reference $U(\boldsymbol{\theta})$ can be reused to estimate expectation values for the target circuit $U(\boldsymbol{\theta}^\prime)$ by reweighting the results with new QPD coefficients.
    
    Let $\mathcal{U}(\boldsymbol\theta)$ and $\mathcal{U}(\boldsymbol\theta^\prime)$ denote the reference and target channels. Assuming a shared set of channels $u_i$, the target channel can be written as
    \begin{equation}
        \label{eq:qpd_reweighting}
        \begin{split}
            &\mathcal{U}(\boldsymbol{\theta}^{\prime})= \sum_{i\in\xi^\prime}a_{i}(\boldsymbol{\theta}^{\prime})
            \, u_{i}\\
            &= \sum_{i\in\xi^\prime\cap\xi} \left\vert\frac{{a_i(\boldsymbol\theta^\prime)}}{a_i(\boldsymbol\theta)}\right\vert \vert a_i(\boldsymbol \theta)\vert \,\mathrm{sgn}(a_{i}(\boldsymbol{\theta}^{\prime}))\, u
            _{i}+ \sum_{i\in\xi^\prime\setminus \xi}a_{i}(\boldsymbol\theta^\prime)\, u_{i}\\
            &= \gamma(\boldsymbol\theta)\sum_{i\in\xi}w_i(\boldsymbol\theta^\prime,\boldsymbol\theta)\,p_{i}(\boldsymbol\theta)\,\mathrm{sgn}(a_{i}(\boldsymbol{\theta}^{\prime})) \,u_{i} + \sum_{i\in\xi^\prime\setminus \xi}a_{i}(\boldsymbol\theta^\prime)\, u_{i}
         \end{split}
    \end{equation}
    where $w_i(\boldsymbol\theta^\prime,\boldsymbol\theta)=\left|{a_i(\boldsymbol\theta^\prime)}/{a_i(\boldsymbol\theta)}\right|$ and $\xi := \{ i : a_i(\boldsymbol{\theta}) \neq 0 \}$ and $\xi' := \{ i : a_i(\boldsymbol{\theta}') \neq 0 \}$. 
    It follows that, if $\xi^\prime \subseteq \xi$, i.e. 
    \begin{equation}
        \forall i,\,  a_{i}(\boldsymbol\theta^{\prime})\neq 0 \Rightarrow a_{i}(\boldsymbol\theta)\neq 0 \,,
    \end{equation}
    then $\mathcal{U}(\boldsymbol{\theta}^\prime)$ can be fully recovered by the reference QPD set $\xi$. For notational simplicity, we omit the explicit parameter dependence and primed quantities refer to the target channel, e.g. $a_i^\prime = a_i(\boldsymbol\theta^\prime)$.
    
    Using Monte Carlo sampling, we construct a reweighted estimator $\hat{\mathcal{U}}(\boldsymbol{\theta}^\prime)$ by reusing samples from the reference distribution. Its expectation value is:
    \begin{equation}
        \label{eq:biased_reweighted_estimator}
        \begin{split}
            \mathbb{E}\left[\hat{\mathcal{U}}^\prime\right]
            = \gamma\sum_{i\in\xi}p_{i}\,w_{i}\,\mathrm{sgn}(a_{i}^{\prime}) \,u_{i}
            =\mathcal{U}^\prime-\sum_{i\in\xi^\prime
        \setminus\xi}a_{i}^{\prime}\,u_{i}\,,
        \end{split}
    \end{equation}
    from which we can conclude that the bias depends on the mismatch between the supports, vanishing when $\xi^\prime \subseteq \xi$. Eq.~\eqref{eq:biased_reweighted_estimator} further shows that, even when $\xi^\prime\not\subset\xi$, we can still reuse a large fraction of the sampled circuits, provided that
    the target and reference QPDs share the same decomposition basis and have
    sufficiently overlapping support.

    When estimating the expectation value of an observable $O$, the above result
    implies that $\hat{O}^\prime:=\mathrm{tr}(O\,\hat{\mathcal{U^\prime}})$ is an unbiased estimator of
    $\mathrm{tr}(O\,\mathcal{U}^\prime)$ and, since the sampling counts $\hat{n}_i$
    follow a multinomial distribution~\cite{evans2000statistical}, its variance is given by

    \begin{equation}
        \label{eq:variance_reweighted_qpd}
        \mathrm{Var}\!\left[\hat{O}^\prime\right]
        =\frac{\gamma }{M}\sum_{i\in\xi}\frac{|a_{i}^{\prime}|^{2}}{|a_{i}|}
        o_i^{2}- \frac{1}{M}\left[\sum_{i\in\xi}a^{\prime}_{i}o_i\right]^{2}\,,
    \end{equation}
    where $o_i:=\mathrm{tr}(O\,u_{i})$.
    This shows that efficient reweighting requires channels with large $|a_i'|$ to be well sampled in the reference distribution.

    When $\xi^{\prime} \subseteq\xi$, the estimator is unbiased and the root-mean-square error reduces to the standard deviation. In this regime, the accuracy of the reweighted estimator depends directly on the variance of the reweighted estimator, which is upper-bounded by $\max_i |o_i|^2\,\chi \gamma^2/M$, where $\chi$ is defined as
    \begin{equation}
        \chi(\boldsymbol\theta, \boldsymbol\theta^{\prime}) = \sum_{i\in\xi}\frac{|a_{i}(\boldsymbol{\theta}^\prime)|^{2}}{\gamma|a_{i}(\boldsymbol{\theta})|}\,.
    \end{equation}
    If $O$ is a Pauli operator, $|o_i| \leq 1$ (as in \eqref{eq:fidelity_estimators} where $o_i = \braket{Z}_i$). The factor $\chi$ thus quantifies the change in variance induced by reweighting relative to the reference QPD.
    
    In particular, $\chi = 1$ when $|a_i| = |a_i'|$, in which case the reweighted estimator has the same variance as the reference QPD; values $\chi > 1$ correspond to an increase in variance, while $\chi < 1$ correspond to a reduction. Thus, $\chi$ identifies regions in parameter space where reweighting remains efficient, with $\chi\lesssim 1$ corresponding to stable regimes.

    A natural expectation is that reweighting remains effective whenever the sampling overhead of the target gate is smaller than that of the reference gate, i.e., when comparing the corresponding $\gamma$ factors. However, the sampling overhead alone does not capture the support of the underlying distributions, which ultimately determines whether reweighting is feasible, as indicated by \eqref{eq:variance_reweighted_qpd}.

\begin{figure}[t]
        \centering
        \includegraphics[width=0.9\linewidth]{
            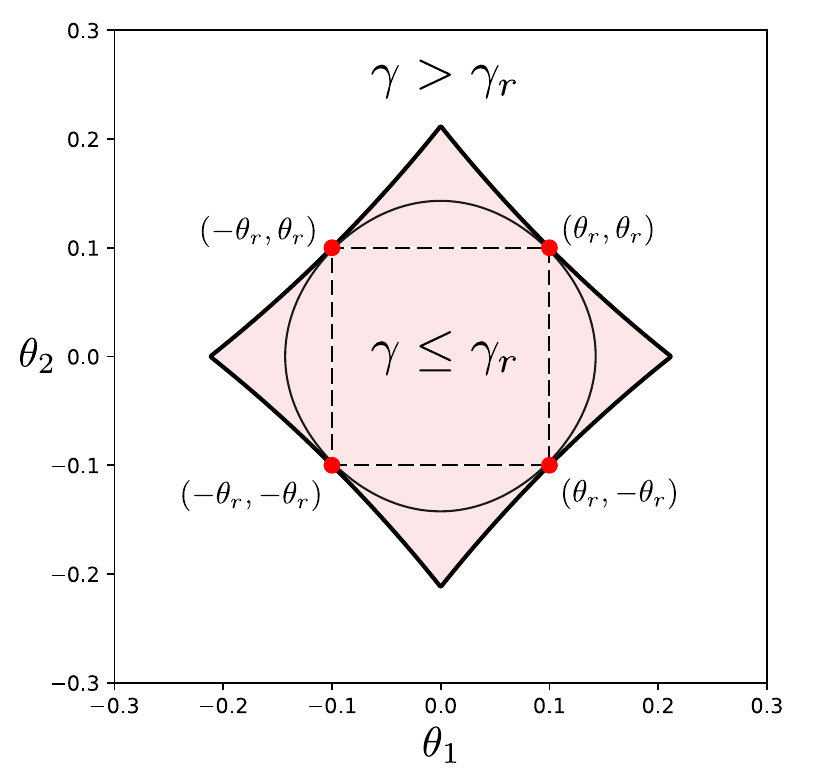
        }
        \caption{Comparison of $\gamma$ and $\chi$ of the QPD for two CR$_Z$ gates, parameterized
        by the pairs $(\theta_{1},\theta_{2})$. The red
        points $(\pm\theta_{r},\pm\theta_{r})$
        represent the reference angles. By symmetry, all pairs $(\pm \theta_r,\pm \theta_r)$
        lead to the same $\gamma_{r} = \gamma(\pm\theta_{r},\pm\theta_{r})$ and
        $\chi$, and, thus, the same resulting plot. The shaded region indicates
        the parameters for which $\gamma(\theta_{1},\theta_{2}) \leq \gamma_{r}$,
        with the boundary given by the points that satisfy the $\gamma$ equality; the
        thin black curve corresponds to the points for which $\chi=1$; and the dashed
        square denotes the perturbation domain $[-\theta_{r},\theta_{r}]^{2}$,
        within which the variance indicator satisfies $\chi \leq 1$.}
        \label{fig:gamma_threshold}
    \end{figure}

    To illustrate this, consider the $\chi$ and $\gamma$ factors of the CR$_Z$ QPD shown in Fig.~\ref{fig:gamma_threshold}, where $\gamma_r$ denotes the reference value. We observe parameter points for which $\gamma < \gamma_r$, yet $\chi>1$ and, therefore, the variance remains large. This highlights that $\chi$ provides a more restrictive and informative criterion, as it accounts for the relative magnitudes of the target and reference coefficients. For this QPD, we further find that parameter values within an $\ell_\infty$-neighborhood of a given reference point can be reliably reweighted.
    
    \subsubsection{Self-normalized importance sampling estimator}
    \label{sec:snisampling}
    In practice, the reweighted estimator $\hat{\mathcal{U}}'$ is biased at finite sample size. 
    Even when $\xi' \subseteq \xi$, the normalization condition 
    $\sum_{i\in\xi} \hat{p}_i w_i = 1$ holds only in expectation, since the weights 
    are random variables. This results in a variance of order $\mathcal{O}(1/M)$, which can become significant when only a limited number of samples is available, as is often the case when cutting many gates.
    
     Self-normalized importance sampling estimators are often employed in such cases as variance reduction methods, yielding a comparatively smaller variance, while remaining asymptotically unbiased~\cite{Owen_MonteCarlo_2013}.
     To construct such an estimator, we normalize the unbiased Monte Carlo estimator~\eqref{eq:qpd_reweighting} with the sum of the reweighted empirical probabilities:
    \begin{equation}
        \hat{\mathcal{U}}_{sn}^\prime
        = \frac{\gamma^\prime \sum_{i\in\xi} \hat{p}_i w_i \mathrm{sgn}(a_i^\prime) u_i}
        {\sum_{i\in\xi} \hat{p}_i w_i}\,.
    \end{equation}
    
    \subsection{Application to the Quantum Geometric Tensor via SPSA}
    \label{sec:qgt_theory}

    Let $\ket{\phi(\boldsymbol \theta)}$ be a quantum state parameterized by $\boldsymbol \theta \in \mathbb{R}^{d}$. The quantum geometric tensor (QGT) is the complex-valued matrix $\mathcal{Q}\in\mathbb{C}^{d \times d}$ with entries~\cite{Gacon_spsa_2021}
    \begin{equation}
        \label{eq:qgt}
        \mathcal{Q_{\mu\nu}}= \braket{\partial_\mu\phi|\partial_\nu\phi}
        - \braket{\partial_\mu\phi|\phi}\!\braket{\phi|\partial_\nu\phi}\, ,
    \end{equation}
    with $\partial_{\mu}:=\frac{\partial}{\partial \theta_{\mu}}$. Its real part, denoted by $\boldsymbol g$, plays a central role in variational quantum algorithms, particularly in variational dynamics and natural-gradient methods.
    
    An alternative expression relates $\boldsymbol g$ to the fidelity $F(\boldsymbol \theta',\boldsymbol \theta ):=|\langle \phi(\boldsymbol \theta')|\phi(\boldsymbol \theta) \rangle |^{2}$ via~\cite{Gacon_spsa_2021}
    \begin{equation}
        \label{eq:qgt_hessian}
        g_{\mu\nu}(\boldsymbol \theta)= \left.-\frac{1}{2}\frac{\partial^{2}}{\partial\theta_{\mu}\partial\theta_{\nu}}
        F(\boldsymbol \theta',\boldsymbol\theta)\right|_{\boldsymbol\theta'=\boldsymbol\theta}\, .
    \end{equation}
    The direct evaluation of this equation requires computing $\mathcal{O}(d^2)$ fidelities. A more efficient approach is provided by the SPSA algorithm~\cite{Gacon_spsa_2021}, which estimates all components simultaneously by sampling random perturbation directions $\boldsymbol{\Delta}_{1}^{(k)}, \boldsymbol{\Delta}_{2}^{(k)} \sim \mathrm{Unif}(\{-1,1\}^d)$ over $K$ iterations. The resulting estimator reads
    \begin{equation}
        \label{eq:qgt_spsa}
        \hat{\boldsymbol{g}}= \frac{1}{K}\sum_{k=1}^{K}\left[-\frac{1}{2}\frac{\delta F^{(k)}}{4h^{2}}
        \phi_{12}^{(k)}\right] \, ,
    \end{equation}
    with
    \begin{equation}
        \phi^{(k)}_{12}= \frac{1}{2}\left(\boldsymbol{\Delta}_{1}^{(k)}\boldsymbol{\Delta}_{2}^{(k)\top}
        +\boldsymbol{\Delta}_{2}^{(k)}\boldsymbol{\Delta}_{1}^{(k)\top}\right)\,,
    \end{equation}
    and
    \begin{equation}
        \delta F^{(k)}= F^{(k)}_{(+,+)}- F^{(k)}_{(+,-)}- F^{(k)}_{(-,+)}+ F^{(k)}_{(-,-)}\,,
    \end{equation}
    where $F^{(k)}_{(\pm,\pm)}:= F(\boldsymbol \theta, \boldsymbol \theta \pm h(\boldsymbol\Delta_{1}^{(k)}\pm \boldsymbol\Delta_{2}^{(k)}))$. The statistical properties of this estimator, including bias and variance scaling, are analyzed in Appendix~\ref{app:spsa_error}.
    
    The dominant cost of this method lies in estimating the fidelity terms. In the following, we show how to compute these efficiently using the QPD-based fidelity estimation and reweighting techniques introduced in Section~\ref{sec:hadamard_qpd}.

    \subsubsection{QGT computation with reweighting}

    Evaluating~\eqref{eq:qgt_spsa} requires $4K$ fidelity computations, which corresponds to a total of $4K N$ circuit runs when the fidelity is evaluated using the standard compute–uncompute method, with $N$ shots per circuit, and twice as many, $8K N$, when using the (uncut) Hadamard test. Moreover, both approaches require circuits with approximately twice the depth of the state-encoding ansatz, which increases the error probability.
    
    To address this overhead, we propose an approach that estimates all $4K$ fidelity terms from a single reference QPD sampling. The method combines the SPSA algorithm with the QPD-based fidelity-estimation and reweighting techniques introduced in this work. As a result, the number of circuit evaluations is reduced to at most $M$, the number of QPD samples.

    The reference fidelity must be chosen such that it is compatible with all fidelity terms appearing in the estimator, allowing them to be obtained through reweighting. Notice that any combination of $\boldsymbol{\Delta}_{1}^{(k)}$ and $\boldsymbol{\Delta}_{2}^{(k)}$ produces a displacement vector whose components lie in $\{-2,0,2\}$. In particular, these vectors are bounded in the $\ell_{\infty}$-norm by the vector with all entries equal to $2$. We therefore choose as reference the fidelity between the states $\ket{\phi(\boldsymbol{\theta})}$ and $\ket{\phi(\boldsymbol{\theta}+2h\mathbf{1})}$, consistent with the analysis of the optimal reweighting region for the CR$_Z$ QPD.
    
    Once the reference fidelity is fixed, its QPD decomposition is sampled to produce a set of circuit instances, which can then be reused to estimate the fidelities appearing in the SPSA estimator through classical reweighting, as discussed in Section~\ref{sec:reweighting}. Note that, although the fidelity estimator is obtained by evaluating the real and imaginary parts of the overlap separately, the corresponding QPD decompositions are identical. Consequently, it suffices to estimate, for example, the real part of the overlap and reuse the same measurement outcomes to reconstruct the imaginary part.
    
    The remaining fidelity terms $F^{(k)}_{(\pm,\pm)}$ are then obtained by applying the reweighting procedure to the sampled circuits. In particular, the real part of the overlap is given by the self-normalized importance sampling estimator
    \begin{equation}
        \hat{R}^\prime= \frac{\gamma^\prime\sum_{i\in\xi}\hat{p}_{i}\,
        w_{i}\hat{{\sigma}}_{i}\braket{Z}_{i}}{\sum_{i\in\xi}\hat{p}_{i}w_{i}}
        \,,
    \end{equation}
    where $\hat{{\sigma}}_{i}$ is defined as in Section~\ref{sec:snisampling}, combining the sign of the coefficient $a_{i}^\prime$ in the target QPD and the measurement outcomes of the reference QPD,
    \begin{equation}
        \hat{{\sigma}}_{i}= \mathrm{sgn}(a_{i}^\prime) \sum_{s\in \mathcal{O}_i}\hat{q}_{s}\,\prod_{m\in s}(-1)^{\delta_{m,1}}\, .
    \end{equation}
    The imaginary part is estimated analogously, with $\braket{Z}_i$ computed using the lookup table described in Appendix~\ref{app:top_exp_value}.

    \begin{figure*}[ht]
        \centering
        \includegraphics[width=\linewidth]{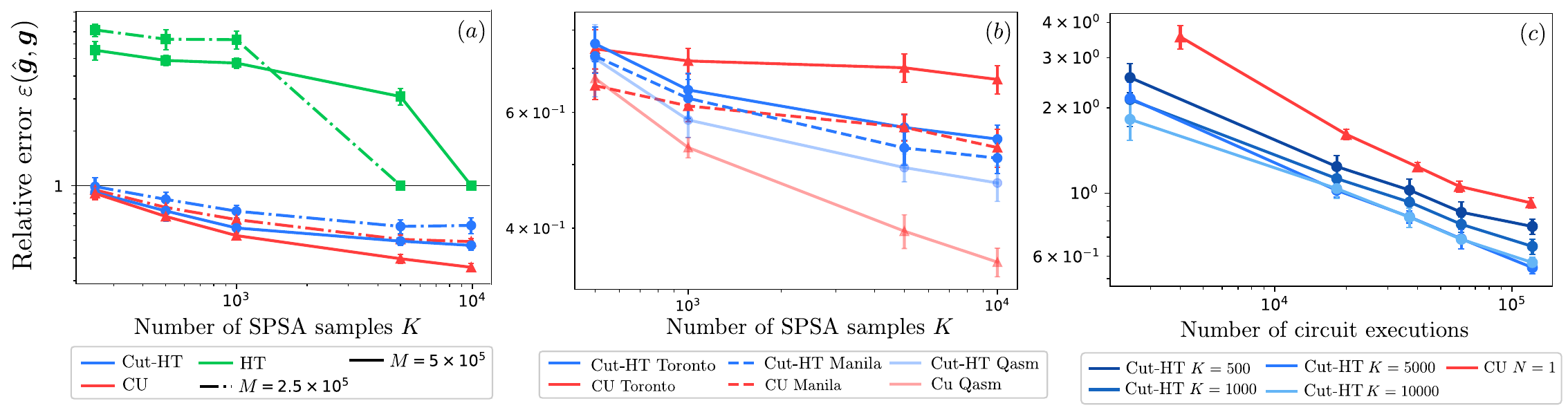}
        \caption{
    Numerical results for the estimation of the real part of the QGT of a 3-qubit system described by the layered ansatz illustrated in Fig.~\ref{fig:efficient_su2} with $L=2$. All data points are averaged over 10 runs. 
    (a) Relative error vs.~the number of SPSA samples $K$ in a noiseless setting. The fidelity is computed using the cut-HT, HT, and CU methods, constrained to the same total number of circuit executions. We set $h=0.1$. 
    (b) Relative error vs.~the number of SPSA samples $K$ on Qiskit is fake Manila and Toronto backends with $M = 5\times10^{5}$ QPD samples. The fidelity is computed using the cut-HT and CU methods, constrained to the same total number of circuit executions. We set $h=0.1$ and compare with the noise-free results obtained with Qiskit's Aer Qasm simulator.
    (c) Relative error vs.~the number of circuit executions on Qiskit's fake Toronto backend. The fidelity is computed using the cut-HT and CU methods. The cut-HT uses $M=\{10^4, 7.5\times10^4,1.5\times10^5,2.5\times10^5,5\times10^5\}$ QPD samples, with each curve corresponding to a fixed number of SPSA samples $K$. For the CU method, $N=1$ shot per circuit is used, while varying $K=\{10^3,5\times10^3,10^4,1.5\times 10^4,3\times 10^4\}$.}
    \label{fig:n3_g_error}
    \end{figure*}
    
    \subsubsection{Numerical results}
    
    We now evaluate the performance of the proposed QGT-estimation method and compare it with the CU approach and both the HT and cut-HT in different noise settings. We simulate a 3-qubit system described by a layered ansatz with $L=2$ layers on different Qiskit 1.2 backends~\cite{Qiskit} and evaluate the accuracy of these methods in terms of the relative error of the estimated QGT, defined with respect to the Frobenius norm:
    \begin{equation}
        \varepsilon(\hat{\boldsymbol{g}},\boldsymbol g)= \frac{\|\hat{\boldsymbol g}-\boldsymbol g\|_{F}}{\|\boldsymbol g\|_{F}}= \frac{\sqrt{\sum_{\mu,\nu=1}^{d}\left(\hat{g}_{\mu\nu}-g_{\mu\nu}\right)^{2}}}{\sqrt{\sum_{\mu,\nu=1}^{d}g_{\mu\nu}^{2}}}
        \, .
    \end{equation}
    
    To provide a fair comparison, we compute the relative error for different numbers of SPSA and QPD samples while constraining the total number of circuit executions to be the same. Specifically, the CU method with $N$ shots per circuit requires $4KN$ circuit executions, whereas the HT requires $8KN$ executions, and the cut-HT requires at most $M$ executions, since it uses a single reference QPD. Consequently, for fixed $N$, the total number of circuit executions for the CU and HT methods grows linearly with $K$, whereas for the cut-HT it remains constant and equal to $M$.
    
    In Fig.~\ref{fig:n3_g_error}$\,$(a), we compare the three methods in a hardware-noise free setting under equal circuit-execution budgets executed on Qiskit's Aer Qasm simulator, which is shot-based but includes no device noise. For the chosen simulation parameters, the CU method achieves the highest accuracy, while the cut Hadamard test performs comparably, particularly when $K$ is small. By contrast, the relative error of the HT remains above $1$, indicating that substantially more circuit executions are required for reliable estimates.
    
    This behavior of the Hadamard test can be understood from its statistical properties. In the low-shot regime, the real and imaginary parts of the overlap tend to take values close to $\pm1$. Since these are squared when computing the fidelity, the estimate can approach $2$, which we cap at $1$. Consequently, the estimated $\hat{\boldsymbol g}$ tends toward an all-zero matrix, yielding a relative error close to $1$.
    
    In the presence of noise, the simplified structure of the circuits used in the cut Hadamard test becomes advantageous. The QPD-based implementation replaces deep entangling circuits with classical post-processing, whereas the CU method requires significantly more CNOT gates due to the doubled circuit depth. To demonstrate this, we estimate the QGT on Qiskit's fake Manila and Toronto backends, which model realistic device noise. As shown in Fig.~\ref{fig:n3_g_error}$\,$(b), the cut Hadamard test outperforms the CU method across different values of $K$, with the advantage increasing for higher CNOT error rates. These results highlight the practical benefits of the QPD approach for near-term devices.
    
    Finally, we compare both methods in terms of circuit executions as a measure of quantum resources on the fake Toronto backend. The HT is omitted, as it is less effective in the previous results. The cut Hadamard test is evaluated for varying $M$, with each curve corresponding to a fixed $K$, whereas the CU method uses $N=1$ (which is optimal, as shown in Fig.~\ref{fig:n3_g_error}$\,$(a)) and varies $K$. From Fig.~\ref{fig:n3_g_error}$\,$(c), we conclude that the cut-HT consistently outperforms the CU method in the considered noisy setting.

    \section{Conclusions \& Outlook}
    \label{sec:conclusions}

    In this work, we developed a sample-efficient reweighting strategy for QPDs
    that enables the reuse of sampled circuit instances across multiple
    related quantum computations. The method
    builds on self-normalized importance sampling and applies whenever the target
    and reference QPDs share the same decomposition basis and have sufficiently
    overlapping support. Within this framework, we analyzed the bias and variance
    of the resulting estimator and identified conditions under which reweighting is reliable.
    
    As a first application, we considered the estimation of fidelities between parameterized
    quantum states. By combining the compressed Hadamard test with QPD circuit
    cutting, we obtained a fidelity-estimation protocol that preserves the structure of the original ansatz and does not require additional qubits
    or increased circuit depth.

    Second, we applied this fidelity reweighting strategy to the estimation of the real part of the QGT via SPSA using a single reference QPD sampling.
   
    We evaluated the relative error of the QGT under a fixed budget
    of circuit executions and across different noise models and observed that, while in noiseless simulations, the compute-uncompute method achieves the highest accuracy, in noisy simulations the cut Hadamard test performs best.
    
    While the proposed approach offers several advantages, its scalability warrants further investigation. The sampling overhead inherent to QPD-based methods grows exponentially with the number of parameters and can be difficult to overcome in general. While reducing the magnitude of the cut angles can alleviate this cost, such reductions may restrict the range of applicability of the method. In this context, reweighting provides a way to better exploit the available samples by enabling their reuse across parameter settings, rather than modifying the underlying scaling. Consequently, the method is most advantageous in regimes involving a large number of target evaluations, such as when many SPSA samples are required for QGT estimation, or when noise levels are sufficiently high that preserving hardware-efficient circuit structures outweighs the sampling overhead.

    We would also like to remark that the classical overhead is also non-negligible, since the weights associated with the sampled circuits must be recomputed for each new parameter set, requiring up to $M\times n_c$ coefficient evaluations for each parameter set. However, these evaluations are mutually independent and can therefore be performed fully in parallel.

    There are several natural directions for future work. One direction is the development
    of optimized, possibly adaptive, reference-selection strategies that choose the
    reference QPD so as to minimize the variance across a family of target channels
    rather than for a single calculation. It would also be interesting to explore
    variance-reduction techniques, e.g. control variates. Beyond
    fidelity estimation and SPSA-based QGT evaluation, the same reweighting framework
    may be useful in a broader class of variational quantum algorithms in which many
    related expectation values must be computed for nearby parameter configurations,
    including variational time evolution, quantum natural-gradient methods, and
    kernel-based quantum machine learning.
    
    Overall, our work shows that quasi-probability decompositions can be used not
    only as a circuit-cutting primitive, but also as a flexible sampling framework
    in which information obtained from one quantum computation can be systematically
    transferred to many others. We expect this perspective to broaden the range of
    practical applications of QPDs and to extend the simulation capabilities of near-term quantum devices.
    
\section*{Acknowledgment}
We acknowledge the use of IBM Quantum services for this work. The views expressed are those of the authors, and do not reflect the official policy or position of IBM or the IBM Quantum team.

\appendices

\section{QPDs of the R$_{ZZ}$ and R$_Z$ gates}\label{app:crz_decomposition}
\begin{figure}[h]
        \centering
        \includegraphics[width=\linewidth]{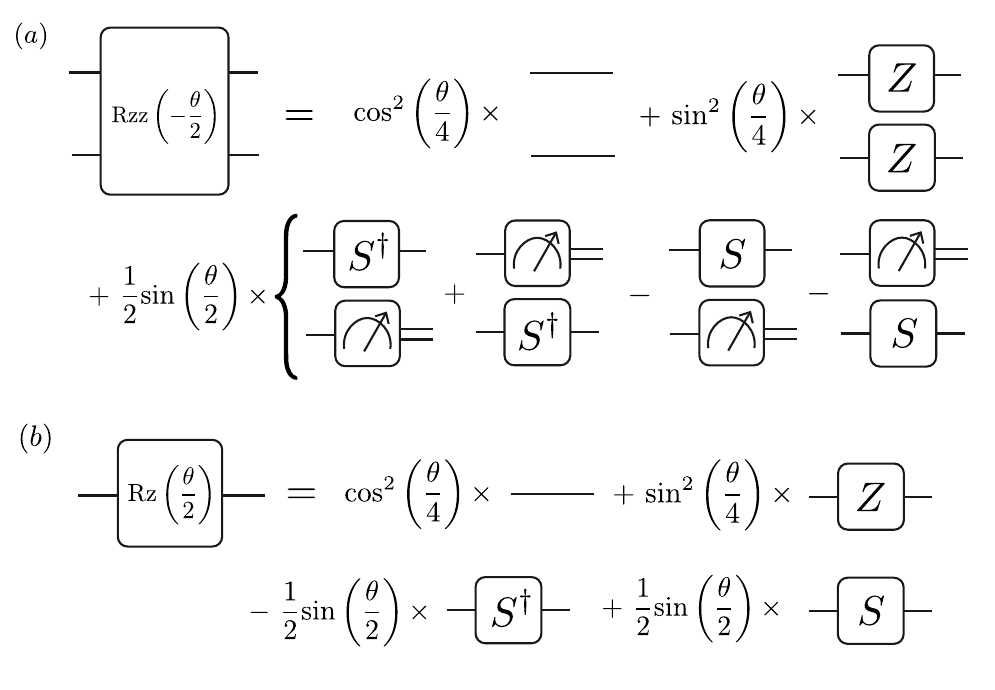}
        \caption{QPDs for (a) the $\mathrm{R}_{ZZ}\left(-{\theta}/{2}\right)$
        and (b) the $\mathrm{R}_Z\left({\theta}/{2}\right)$ gates. Empty wires
        represent the identity operation, $S=\sqrt{Z}$, and measurements are in the
        computational basis.}
        \label{fig:R_ZZ_rz_qpd}
    \end{figure}

The QPDs of the R$_{ZZ}$ and R$_Z$ gates used in the derivation of \eqref{eq:qpd_crz} are shown in Fig.~\ref{fig:R_ZZ_rz_qpd}.

\section{Variance of the fidelity estimated via QPD}
\label{app:variance_fidelity_qpd}

Let $\hat{R}$ and $\hat{J}$ be estimators of the real and imaginary parts of the overlap between two quantum states, obtained by sampling $M$ circuits from the QPD set, and let $\hat{F}$ denote the empirical fidelity defined as $\hat{F}= \hat{R}^{2}+\hat{J}^{2}$.
Since $\hat{R}$ and $\hat{J}$ differ only in $\braket{Z}_{i}$, we analyze $\hat{R}$ (the same results apply to $\hat{J}$). The estimator reads
\begin{equation}
    \hat{R}= \gamma \sum_{i\in\xi}\hat{p}_{i}\,\mathrm{sgn}(a_{i})
    \langle Z\rangle_{i}(2\hat{q}_{i}-1)\,,
\end{equation}
where $\hat{p}_{i}=\hat{n}_{i}/M$ and $\hat{q}_{i}=\hat{n}_{i}^{+}/\hat{n}_{i}$, with $\hat{n}_{i}^{+}$ the number of measurements with outcome $0$.
Although $\hat{p}_{i}$ and $\hat{q}_{i}$ both depend on $\hat{n}_i$, their product satisfies
\begin{equation}
    \mathbb{E}\left[\hat{p}_{i}\hat{q}_{i}\right]= \mathbb{E}\left[\frac{\hat{n}_{i}}{M}\sum_{m=1}^{\hat{n}_{i}}\delta_{\{\hat{Q}_{m}=0\}}\right]=p_{i}q_{i}\,,
\end{equation}
where $\hat{Q}_{m}\in\{0,1\}$ is the measurement outcome of shot $m$ and in the second line we used that $\hat{q}_{i}$ is a Bernoulli distributed random variable with parameter $q_{i}$, which is the exact probability of obtaining the outcome $+1$ from circuit $i$.

The first moment of $\hat{R}$ is then
\begin{equation}
    \mathbb{E}\left[\hat{R}\right]
    = \gamma \sum_{i\in\xi}p_{i}\,\mathrm{sgn}(a_{i})\braket{Z}_{i}(2q_{i}-1)
    = R\,,
\end{equation}
and we conclude that $\hat{R}$ is unbiased. Similarly, $\mathbb{E}[\hat{J}]=J$.

Using standard results for multinomial sampling~\cite{evans2000statistical}, we obtain that
\begin{equation}
\begin{split}
   &\mathbb{E}\left[\hat{p}_{i}\hat{p}_{j}(2\hat{q}_{i}-1)(2\hat{q}_{j}-1)\right]
    =\\
    &=\left(1-\frac{1}{M}\right)p_{i}(2q_{i}-1)p_{j}(2q_{j}-1)
    + \frac{1}{M}\delta_{i,j}p_{i}\,.
\end{split}
\end{equation}
Substituting into $\mathbb{E}[\hat{R}^2]$ gives
\begin{equation}
    \mathbb{E}\left[\hat{R}^{2}\right]
    = \left(1-\frac{1}{M}\right)R^{2}
    + \frac{\gamma^{2}}{M}\sum_{i\in\xi}p_{i}\braket{Z}_{i}^{2}\,,
\end{equation}
and the variance reads as
\begin{equation}
    \mathrm{Var}\left[\hat{R}\right]
    = \frac{\gamma^{2}}{M}\sum_{i\in\xi}p_{i}\langle Z \rangle_{i}^{2}
    - \frac{R^{2}}{M}\,.
\end{equation}
An analogous expression holds for $\hat{J}$.

\subsection{Bias of the fidelity estimator}

The expectation of $\hat{F}$ is
\begin{equation}
    \mathbb{E}\left[\hat{F}\right]
    = \left(1-\frac{1}{M}\right)F + \frac{\gamma^{2}}{M}\,,
\end{equation}
where we used that $\langle Z\rangle_i^{\mathrm{re/im}}\in\{-1,0,1\}$ and do not simultaneously take value $1$ (see Appendix~\ref{app:top_exp_value}). The unbiased estimator is therefore
\begin{equation}
    \label{eq:qpd_fidelity_bias}
    \hat{F}^\star = \frac{M\hat{F}-\gamma^2}{M-1}\,.
\end{equation}

\subsection{Mean-squared error of the fidelity estimator}

Assuming $\hat{R}$ and $\hat{J}$ are computed from independent samples,
\begin{equation}
    \mathrm{MSE}[\hat{F}]
    = \mathbb{E}\left[(\hat{R}^{2}-R^{2})^{2}\right]
    + \mathbb{E}\left[(\hat{J}^{2}-J^{2})^{2}\right]
    + 2\,\mathrm{Var}[\hat{R}]\,\mathrm{Var}[\hat{J}]\,.
\end{equation}

For large $M$, we define $\delta R=\hat{R}-R$ and $\delta J=\hat{J}-J$, and expanding to leading order yields,
\begin{equation}
    \mathbb{E}\left[(\hat{R}^{2}-R^{2})^{2}\right]
    \approx 4R^{2}\,\mathrm{Var}[\hat{R}]\,,
\end{equation}
and similarly for $\hat{J}$. Hence,
\begin{equation}
    \mathrm{MSE}[\hat{F}]
    \approx 4R^{2}\mathrm{Var}[\hat{R}]
    + 4J^{2}\mathrm{Var}[\hat{J}]
    = 4F\,\mathcal{O}\!\left({\gamma^{2}}/{M}\right)\,.
\end{equation}

If instead the same QPD samples are reused for both estimators, $\hat{R}$ and $\hat{J}$ are generally correlated, and additional cross terms contribute to the mean-squared error. Nevertheless, the corresponding covariances are bounded by the variances of $\hat{R}$ and $\hat{J}$, and thus have the same $\mathcal{O}(\gamma^{2}/M)$ scaling, affecting only its prefactor.

\section{Exact estimation of the ancilla circuit with no matrix
        multiplications}
\label{app:top_exp_value}
 For a circuit with $d$ unique parameters, the ancilla circuit consists of only $d+2$ or $d+3$ single-qubit operations, depending on whether it corresponds to the real or imaginary Hadamard test, and can be simulated efficiently by performing matrix multiplications. Here, however, we present an alternative strategy that computes the expectation value exactly by tracking the sampled channels, applicable to the QPD of \eqref{eq:qpd_crz}.

First, note that if the state before the final Hadamard gate $H$ is $\ket{0}$ or $\ket{1}$, then $H$ prepares an equal superposition and $\braket{Z}=0$; conversely, if the qubit remains in $\ket{0}$ or $\ket{1}$ at measurement, then $\braket{Z}=1$ or $-1$, respectively. For the channels in this QPD, these are the only possible outcomes.

Next, consider the effect of mid-circuit measurements, which are all performed in the $Z$ basis. If the state is in an equal superposition before the measurement, it collapses to $\ket{0}$ or $\ket{1}$ with equal probability; if it is already in $\ket{0}$ or $\ket{1}$, it remains unchanged. The remaining channels ($S$, $S^{\dagger}$, and $Z$) are diagonal in the computational basis and only affect the global phase of the state. Therefore, if a mid-circuit measurement occurs, the state before the final Hadamard gate is $\ket{0}$ or $\ket{1}$, and thus $\braket{Z}=0$.

In the absence of mid-circuit measurements, the ancilla state can be written as $HZ^{a}S^{b}[S^{\dagger}]^{c}H\ket{0}$, where $a,b,c$ are non-negative integers corresponding to the number of sampled $Z$, $S$, and $S^{\dagger}$ gates modulo 2, 4, and 4, respectively. Since these gates are diagonal and mutually commute, their combined action is
\begin{equation}
Z^a S^b [S^{\dagger}]^c
=
\operatorname{diag}
\left(1,e^{i\pi p/2}\right),\, \textrm{where}\, p=(2a+b-c)\bmod 4\,.
\end{equation}
After the initial Hadamard gate, the ancilla state is therefore
\begin{equation}
\braket{Z}
=
\begin{cases}
1, & p=0, \\[3pt]
0, & p=1 \text{ or } p=3, \\[3pt]
-1, & p=2.
\end{cases}
\end{equation}

Thus, it suffices to track the number of sampled $Z$, $S$, and $S^{\dagger}$ gates modulo 2 or 4. This can be done in $\mathcal{O}(d)$ time, without performing any matrix multiplications.

\section{Error analysis of the QGT estimator via SPSA}
\label{app:spsa_error}

We analyze the statistical error of the SPSA estimator $\hat{ \boldsymbol g}$ introduced in Sec.~\ref{sec:qgt_theory}. 

When both states are described by the same ansatz, differing only by a small perturbation $\boldsymbol{\delta}=\{\pm h\}^d$, e.g., $\ket{\phi(\boldsymbol{\theta})}=U(\boldsymbol{\theta})\ket{0}$ and $\ket{\phi(\boldsymbol{\theta}+\boldsymbol{\delta})}=U(\boldsymbol{\theta}+\boldsymbol{\delta})\ket{0}$, the fidelity can be expanded as
\begin{equation}
    \label{eq:fidelity_expanded}
    F(\boldsymbol{\theta},\boldsymbol{\theta}+\boldsymbol{\delta}) = 1-
    \boldsymbol{\delta}^{\top}\boldsymbol{g}\boldsymbol{\delta}+ \mathcal{O}\!\left(h^{3}\right)\,,
\end{equation}
where $\boldsymbol{g}$ is the real part of the QGT. Using~\eqref{eq:fidelity_expanded} and the symmetry of the QGT,
we obtain
\begin{equation}
    \delta F^{(k)}=-8h^{2}\left[ \boldsymbol{\Delta}_{1}^{(k)}\right]^{\top} \boldsymbol{g}
     \boldsymbol{\Delta}_{2}^{(k)}+ \mathcal{O}\left(d^{2}h^{4} + \varepsilon_F\right) \,,
\end{equation}
including the asymptotic scaling $\varepsilon_F$ of the fidelity estimator RMSE, derived in Appendix~\ref{app:variance_fidelity_qpd}.

Using standard moment identities for independent Rademacher variables, it follows that the expectation value over the $K$ independently sampled SPSA directions is
\begin{equation}
    \label{eq:qgt_bias}
    \mathbb{E}\left[\hat{g}_{ij}\right]
    = g_{ij}+ \mathcal{O}\left(d^{2}h^{2}+ \frac{\varepsilon_{F}}{h^{2}}\right)\,,
\end{equation}
and, thus, the bias scales as
$\mathcal{O}\left(d^{2}h^{2}+{\varepsilon_{F}}/{h^{2}}\right)$. Using independence across SPSA samples and Rademacher-moment identities, it follows that
\begin{equation}
    \begin{split}
        \mathrm{Var}\left[\hat{g}_{ij}\right]
        &= \frac{1}{K}\left[\frac{1}{2}\mathrm{tr}\left(\boldsymbol{g}^{2}\right)
        \left(1+\delta_{i,j}\right) - 2g_{ii}^{2}\delta_{i,j}+ g_{ii}g_{jj}
        \right]  \\
        &\quad + \mathcal{O}\left(\frac{d^{2}\varepsilon_{F}}{K}
        +\frac{\varepsilon_{F}^{2}}{h^{4}K}
        +\frac{d^{2}h^{2}}{K}
        +\frac{\varepsilon_{F}}{h^{2}K}\right)\,.
    \end{split}
\end{equation}

Combining the bias and variance contributions, the squared Frobenius-norm RMSE is   
\begin{small}
\begin{equation}
    \label{eq:rmse_qgt}
    \begin{split}
        &\sqrt{\mathbb{E}\left[||\hat{\boldsymbol{g}}-\boldsymbol{g}||^{2}_{F}\right]}
        =\sqrt{\frac{1}{K}\left[\mathrm{tr}\left( \boldsymbol{g}^{2}\right)
        \frac{d(d+1)}{2} + \mathrm{tr}\left( \boldsymbol{g}\right)^2-2\sum_{i=1}^dg^2_{ii}\right]} \\
        &\quad + \mathcal{O}\left(
        \sqrt{
        \frac{d^{4}\varepsilon_{F}}{K}
        +\frac{d^{2}\varepsilon_{F}^{2}}{h^{4}K}
        +\frac{d^{4}h^{2}}{K}
        +\frac{d^{2}\varepsilon_{F}}{h^{2}K}
        + d^{2}\left(d^{2}h^{2}+\frac{\varepsilon_{F}}{h^{2}}\right)^{2}}
        \right)\,.
    \end{split}
\end{equation}
\end{small}

\bibliographystyle{ieeetr}
\bibliography{bibliography}
\end{document}